\documentclass[sigconf, nonacm]{acmart}

\newcommand{\design}{TRINE}

\usepackage{hyperref}
\usepackage{amsmath,amsfonts}
\usepackage{algorithm,algpseudocode}
\algnewcommand\algorithmicforeach{\textbf{for each}}
\algdef{S}[FOR]{ForEach}[1]{\algorithmicforeach\ #1\ \algorithmicdo}
\usepackage{listings}
\usepackage{array}
\usepackage{graphicx}
\usepackage{textcomp}
\usepackage{xcolor}
\usepackage{orcidlink}
\usepackage{svg}
\usepackage{enumitem}
    
\usepackage{multirow}
\usepackage{makecell}
\usepackage{booktabs}

\usepackage[table]{xcolor}
\usepackage{colortbl}
\definecolor{lightblue}{RGB}{250,250,250}

\usepackage{siunitx}
\usepackage{pgfplots, pgfplotstable}
\pgfplotsset{compat=1.16}
\usepackage{tikz}
\tikzset{axis background/.style={inner sep=0, outer sep=0}}
\tikzset{every axis label/.style={inner sep=0, outer sep=0}}
\definecolor{color0}{HTML}{1C6E8C}
\definecolor{color1}{HTML}{EE7674}
\pgfplotscreateplotcyclelist{colors}{
{fill=color0},
{fill=color1}
}

\usepackage[font=footnotesize]{caption}
\usepackage{verbatim}
\usepgfplotslibrary{external}
\newcommand\resetstackedplots{
\makeatletter
\pgfplots@stacked@isfirstplottrue
\makeatother
\addplot [forget plot,draw=none] coordinates{(1,0) (2,0) (3,0)};
}

\begin{document}

\title{\design: A Token-Aware, Runtime-Adaptive FPGA Inference Engine for Multimodal AI}
\author{Hyunwoo Oh$^1$, Hanning Chen$^1$, Sanggeon Yun$^1$, Yang Ni$^2$, Suyeon Jang$^1$,\\ Behnam Khaleghi$^3$, Fei Wen$^4$, and Mohsen Imani$^1$}
\affiliation{\institution{$^1$University of California, Irvine, $^2$Purdue University Northwest, $^3$Qualcomm, $^4$Samsung}\country{}}
\email{{hyunwooo, m.imani}@uci.edu}

\begin{abstract}
Multimodal stacks that mix ViTs, CNNs, GNNs, and transformer NLP strain embedded platforms because their compute/memory patterns diverge and hard real-time targets leave little slack. \design{} is a single-bitstream FPGA accelerator and compiler that executes end-to-end multimodal inference without reconfiguration. Layers are unified as DDMM/SDDMM/SpMM and mapped to a mode-switchable engine that toggles at runtime among weight/output-stationary systolic, 1×CS SIMD, and a routable adder tree (RADT) on a shared PE array. A width-matched, two-stage top-k unit enables in-stream token pruning, while dependency-aware layer offloading (DALO) overlaps independent kernels across reconfigurable processing units to sustain utilization. Evaluated on Alveo U50 and ZCU104, \design{} reduces latency by up to 22.57× vs. RTX 4090 and 6.86× vs. Jetson Orin Nano at 20–21 W; token pruning alone yields up to 7.8× on ViT-heavy pipelines, and DALO contributes up to 79\% throughput improvement. With int8 quantization, accuracy drops remain $<$2.5\% across representative tasks, delivering state-of-the-art latency and energy efficiency for unified vision, language, and graph workloads—in one bitstream.
\end{abstract}
\maketitle

\vspace{-1pt}

\vspace{-2mm}
\section{Introduction}\label{sec:intro}
Multimodal artificial intelligence (AI) is reshaping computer vision (CV), language understanding, and relational reasoning by jointly modeling images, text, and graphs. Systems that combine Vision Transformers (ViTs), Graph Neural Networks (GNNs), Convolutional Neural Networks (CNNs), and transformer-based Natural Language Processing (NLP) have enabled text-conditioned detection and vision–language grounding \cite{mdetr, clip1} as well as graph-augmented cognition \cite{missiongnn}. Yet these gains come with heterogeneous compute and memory behaviors that complicate utilization, workload balance, and real-time inference on embedded platforms.

ViTs are a particular pain point: fixed-length token processing and attention-heavy blocks inflate the cost of both feed-forward networks and attention matrices, especially in CV pipelines with many tokens and moderate embedding widths \cite{mdetr, vit_ffn_load, vit_big}. Token-pruning methods reduce this cost by discarding low-importance tokens at inference time \cite{vit_token_pruning_0, vit_token_pruning_1, dynamicvit, evovit, spvit}. However, on GPUs the resulting irregular sparsity often degrades utilization, so measured speedups fall short of theoretical potential \cite{spvit, heatvit, vitcod}. Prior hardware efforts have accelerated isolated pieces—e.g., ViT-only or NLP-only pruning—or targeted a subset of kernels, leaving end-to-end multimodal pipelines underserved \cite{heatvit, vitcod, hw_dynamic_pruning, hw_nlp_pruning0, hw_nlp_pruning1}.

FPGAs are a natural fit for such heterogeneity, but frequent bitstream swapping introduces prohibitive reconfiguration delays and complicates deployment \cite{reconfig_latency}. While some designs support multiple workloads \cite{hw_cnn_gnn, hw_cnn_gnn_vit}, a unified solution that adapts at runtime across the full diversity of multimodal AI remains elusive. Moreover, replicating separate dense and sparse engines inflates area and complicates timing closure, often depressing maximum clock frequency. What is missing is a single bitstream that (i) executes ViT/CNN/GNN/NLP end-to-end, (ii) embraces dynamic token sparsity, and (iii) sustains high utilization without reconfiguration.

This work takes the view that multimodal layers can be expressed as three matrix kernels—dense–dense (DDMM), sampled dense–dense (SDDMM), and sparse (SpMM)—provided the hardware can change its dataflow at runtime via switchable interconnects and per-PE operation modes. This abstraction lets a single PE array serve all kernels by switching interconnects and per-PE ops at runtime rather than instantiating duplicated datapaths. We therefore introduce \design{}, an FPGA-based accelerator and compiler that map DDMM/SDDMM/SpMM onto a shared datapath with a mode-switchable engine capable of weight- and output-stationary systolic execution, $1{\times}C_S$ SIMD, and a routable adder tree (RADT) for highly sparse reductions. A width-matched, two-stage top-$k$ unit performs in-stream token pruning so attention scores are filtered as they are produced, avoiding oversized global sorters and off-chip detours. Above the engine, dependency-aware layer offloading (DALO) overlaps independent kernels across multiple reconfigurable processing units (RPUs). RADT and DALO act at complementary levels—RADT handles intra-kernel sparsity and reductions, DALO exposes inter-kernel concurrency—enabling runtime adaptation within a single bitstream without sacrificing efficiency.

\begin{itemize}[leftmargin=*,nosep]
  \item \design{} unifies ViT/CNN/GNN/NLP layers as DDMM/SDDMM/SpMM and executes them end-to-end on one bitstream without reconfiguration.
  \item A shared PE array switches among systolic (WS/OS), $1{\times}C_S$ SIMD, and RADT via fine-grained routing, matching kernel structure and sparsity without duplicating engines.
  \item In-stream top-$k$ pruning is coupled with RPU-level scheduling: the array-width-matched sorter prunes tokens in flight, and DALO overlaps independent kernels to raise utilization on multi-RPU fabrics.
\end{itemize}

We evaluate \design{} on Xilinx Alveo U50 and ZCU104. Across TinyCLIP \cite{tinyclip}, MDETR \cite{mdetr}, and MissionGNN \cite{missiongnn} workloads, \design{} reduces latency by up to $22.57\times$ than RTX~4090 and $6.86\times$ versus a Jetson Orin Nano while operating at $\sim$20–21\,W; pruning alone yields up to $7.8\times$ speedup on ViT-heavy cases, and DALO improves throughput by $79\%$. To our knowledge, this is the first work to provide state-of-the-art latency and energy efficiency across vision, language, and graph workloads within a single bitstream.

\section{Related Works and Motivations}\label{sec:rel_works}

\subsection{Multimodal AI Models}
Multimodal systems combine ViTs, CNNs, GNNs, and transformer NLP to align and reason over images, text, and graphs. MDETR fuses CNN/ViT with language for text-conditioned detection \cite{mdetr}; CLIP/TinyCLIP learn vision–language embeddings for zero-shot tasks \cite{clip0, clip1, tinyclip}; ImageBind extends to additional modalities \cite{imagebind}. On the graph side, MissionGNN converts visual evidence into a knowledge graph processed by a GNN \cite{missiongnn}; ML-CGN mixes CNNs and GNNs for scene/relational understanding \cite{ml_gcn}; TaskCLIP targets fine-grained segmentation \cite{taskclip}. Despite this progress, the workload heterogeneity across NLP/GNN/CNN/ViT complicates efficient acceleration.

\subsection{Token Pruning for ViTs}
ViT cost scales with tokens and attention. Dynamic pruning keeps only the most informative tokens using top-$k$ selection \cite{dynamicvit, evovit}, but on CPUs/GPUs the induced irregular sparsity limits realized speedups. SPViT replaces top-$k$ with Gumbel-Softmax for differentiable selection \cite{spvit}, which introduces nonlinear kernels that are also unfriendly to general-purpose hardware. These trends motivate specialized support for pruning that reduces compute \emph{and} preserves utilization.

\subsection{Domain-Specific Accelerators}
\subsubsection{Hardware Pruning}
Accelerators for sparse kernels (SpMM/SDDMM) and pruned transformers improve throughput but are typically model- or kernel-specific \cite{hw_spmm_sddmm, hw_nlp_pruning0}. Token-pruning hardware exists—e.g., threshold-based engines for NLP \cite{hw_nlp_pruning1}, compile-time pruning for ViTs \cite{vitcod}, and run-time Gumbel-Softmax pruning \cite{heatvit}—yet each targets a subset of models or fixes sparsity patterns, limiting generality.

\subsubsection{Multimodality-Aware Accelerators}
Mode-switchable FPGA designs covering multiple modalities have emerged: CNN+GNN \cite{hw_cnn_gnn} and extensions that add ViT support \cite{hw_cnn_gnn_vit}. However, coverage remains incomplete (not all four modalities) and most lack hardware-accelerated \emph{run-time} token pruning, leaving end-to-end multimodal pipelines under-optimized.

\subsection{Motivations}
As summarized in \autoref{tab:related_works}, prior work either narrows to a few modalities or omits run-time token pruning essential for ViT-heavy pipelines. We aim for a single-bitstream framework that spans ViT/GNN/CNN/NLP and provides scalable, hardware top-$k$ pruning so emerging pruning strategies can be executed in real time within unified multimodal graphs.

\begin{table}[tb!]
\centering
\caption{FPGA accelerators vs. supported workloads, sparse-kernel support, and token-pruning capability.}
\vspace{-2mm}
\label{tab:related_works}
\resizebox{\linewidth}{!}{%
\begin{tabular}{l|c|c|c}
\toprule
\multirow{2}{*}{\textbf{Work}}      & \textbf{AI}           & \textbf{Sparse}   & \textbf{Token}            \\ 
                                    & \textbf{Workload}     & \textbf{Kernel}   & \textbf{Pruning}          \\ \hline \hline
This work                           & ViT+GNN+CNN+NLP       & Yes               & Run-time                  \\ \hline
\cite{hw_cnn_gnn_vit}               & CNN+GNN or ViT        & Yes               & No                        \\ \hline
\cite{hw_cnn_gnn}                   & CNN+GNN               & No                & No                        \\ \hline
\cite{heatvit}                      & ViT                   & No                & Run-time$^\mathrm{*}$     \\ \hline
\cite{vitcod}                       & ViT                   & No                & Compile-time$^\mathrm{*}$ \\ \hline
\cite{hw_nlp_pruning1}              & NLP                   & Yes               & Compile-time$^\mathrm{*}$ \\ \hline
\cite{hw_nlp_pruning0}              & General MM            & Yes               & No                        \\ \hline
\cite{hw_spmm_sddmm}                & General MM            & Yes               & No                        \\
\bottomrule
\multicolumn{4}{l}{\small$^\mathrm{*}$Compile/run-time denotes when the sparsity pattern is determined.} \\
\end{tabular}
}
\vspace{-4mm}
\end{table}

\begin{figure*}[tb]
\centering
\includegraphics[width=\linewidth]{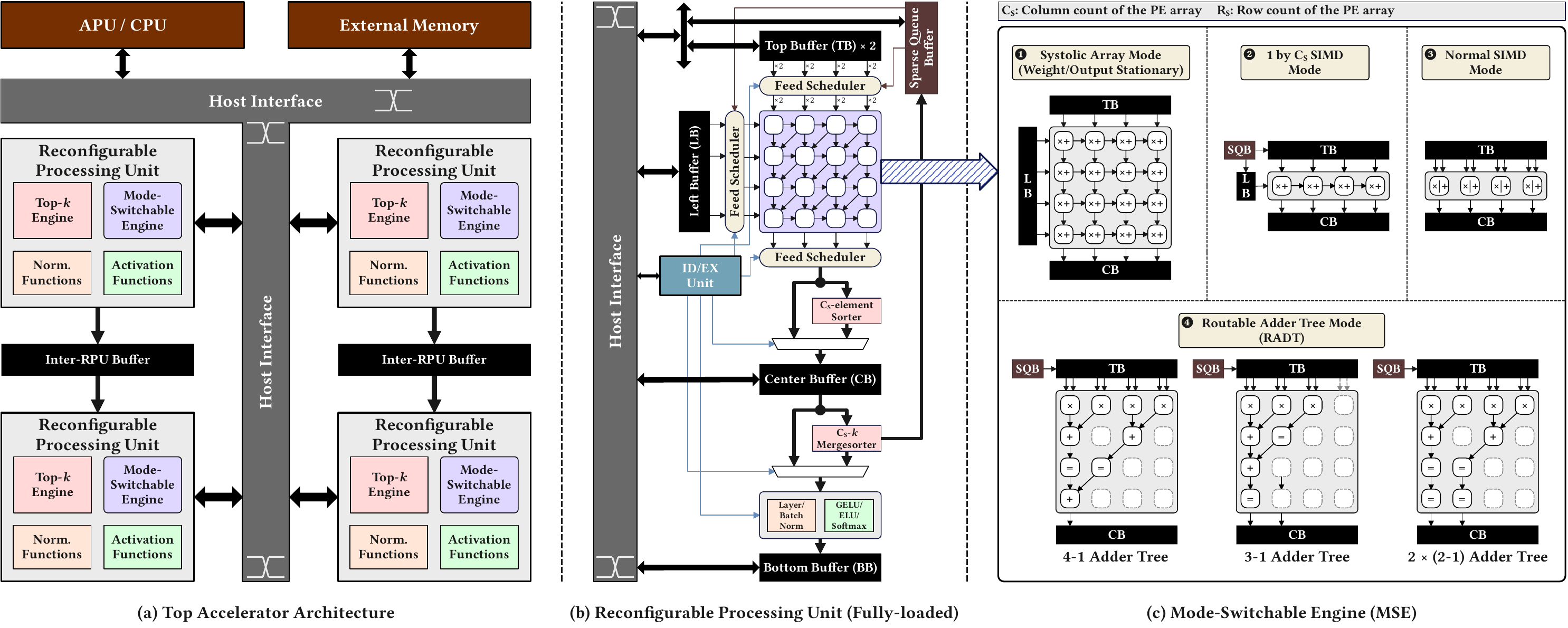}
\caption{
\textbf{\design{} overview and mode-switchable engine (MSE).}
(\textbf{a}) Accelerator with an RPU grid and local inter-RPU buffers to localize traffic and pipeline inter-tile exchange. 
(\textbf{b}) Each RPU integrates a shared-datapath MSE, a width-matched two-stage top-$k$, compact nonlinear units, and a lightweight feed scheduler. 
(\textbf{c}) One PE array time-shares four dataflows via small interconnect muxes and per-PE op control:
(1) Systolic (WS/OS) for dense DDMM;
(2) $1{\times}C_S$ SIMD for moderately sparse SDDMM/SpMM;
(3) RADT for highly sparse/irregular reductions;
(4) normal SIMD for element-wise ops.
Selection policy (runtime): DDMM$\rightarrow$WS/OS (WS for small-token/high weight reuse; OS for wide feature maps/many tokens). 
SDDMM/SpMM$\rightarrow$$1{\times}C_S$ when active operands per row/col $\lesssim C_S$ and fairly uniform; switch to RADT as sparsity grows or degree skews.
Feed scheduling provides systolic delay insertion and sparsity-aware indexed reads without host-side packing.
}
\vspace{-4mm}
\label{fig:hw_arch}
\end{figure*}

\section{Hardware Architecture}\label{sec:sys_arch}
\design{} unifies ViT, GNN, NLP, and CNN inference on a single bitstream by combining a grid of reconfigurable processing units (RPUs), a mode-switchable compute engine (MSE) that switches dataflows on one PE array, and an in-stream, width-matched top-$k$ unit for run-time token pruning (Fig.~\ref{fig:hw_arch}).

\subsection{System Overview}
As shown in Fig.~\ref{fig:hw_arch}(a), an $r{\times}c$ RPU grid with local inter-RPU buffers minimizes off-chip traffic and enables pipelined inter-tile exchange. A host APU/CPU writes compact control blocks (mode, tiling, routing masks) and uses AXI DMA for bulk I/O. Mode changes drain the in-flight pipeline and update a few registers; since kernels run for thousands of cycles, the overhead is negligible in wall time. This separation of control and dataflow allows per-kernel adaptation at runtime without any bitstream reconfiguration.

\subsection{Mode-Switchable Engine (MSE)}
Within each RPU in Fig.~\ref{fig:hw_arch}(b), the MSE is a single PE array that time-shares multiple dataflows using small interconnect multiplexers and per-PE operation control. Each PE provides west/north inputs, a partial-sum path, a tiny register file for reuse, and a three-function ALU (MAC/ADD/PASS). One multiplexer steers the $\mathsf{B}$ stream to realize output- or weight-stationary wavefronts; another selects the PE operation (multiply–accumulate, reduction for RADT, or pass-through). Lightweight row/column broadcasts and a short cross-row tap enable SIMD and tree reductions without duplicating engines or introducing long global wires. Consolidating dense and sparse dataflows into one array avoids a second large macro and its routing pressure, which helps timing closure and keeps $F_{\max}$ stable when pruning is enabled.

The four execution modes are summarized in Fig.~\ref{fig:hw_arch}(c):
\begin{itemize}[leftmargin=*,nosep]
  \item \textbf{Systolic (WS/OS).} Dense DDMM (conv-as-GEMM, MLP, full attention) with reuse balanced between weights (WS, small tokens/large reuse) and outputs (OS, wide feature maps/many tokens).
  \item \boldmath$\mathbf{1{\times}C_S}$ \textbf{SIMD}. A single active row of width $C_S$ (\emph{array width in columns}) issues only scheduled nonzeros; effective for SDDMM/SpMM at moderate sparsity or bounded degrees (GNN).
  \item \textbf{RADT}. The same PEs form a programmable multi-stage reduction tree; active products are injected where they occur and accumulated along routed paths, preferred under high or skewed sparsity.
  \item \textbf{Normal SIMD}. Element-wise work such as bias, activation, and edge functions.
\end{itemize}

Mode selection uses kernel shape $(N,M,K)$, observed token/edge sparsity $p$, and array width $C_S$: dense layers map to WS/OS; pruned attention and message passing use $1{\times}C_S$ when activity per row is near $C_S$ and fairly uniform, and switch to RADT when activity is much smaller or heavily skewed. In our builds, sharing one array instead of duplicating dense/sparse macros reduced logic/routing by roughly 20–30\% while maintaining $F_{\max}$.

\subsection{Top-$k$ Engine (In-Stream Pruning)}
\begin{figure}[tb!]
\centering
\includegraphics[width=\linewidth]{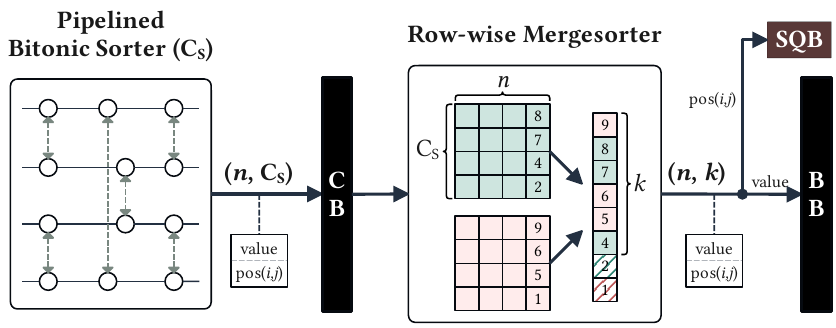}
\caption{Two-stage, width-matched top-$k$. A fixed-width (up to $C_S$) pipelined bitonic stage matches array width; a lightweight $C_S{\rightarrow}k$ merge completes selection. Values/indices stream from MSE into a center buffer (CB) and sparse queue buffer (SQB), avoiding off-chip detours and scaling better than single large bitonic networks.}
\vspace{-4mm}
\label{fig:top_k_engine}
\end{figure}
Token and score selection happens immediately after compute. The unit is split into a fixed-width (up to $C_S$) pipelined bitonic stage followed by a small $C_S{\rightarrow}k$ mergesorter. Matching the first stage to the array’s output width (Fig.~\ref{fig:hw_arch}(c)) caps the number of compare-and-swap units; the merge completes selection with little extra logic. Values and indices stream through a center buffer and sparse queue buffer, so pruning neither stalls the pipeline nor detours through DRAM. Compared with a single large bitonic network whose area grows as $O(n\log^2 n)$, this width-matched split preserves near-streaming throughput at substantially lower area, and the microarchitecture can be swapped (dual-layer or FLiMS) when devices are tight.

\begin{figure}[tb!]
\centering
\includegraphics[width=\linewidth]{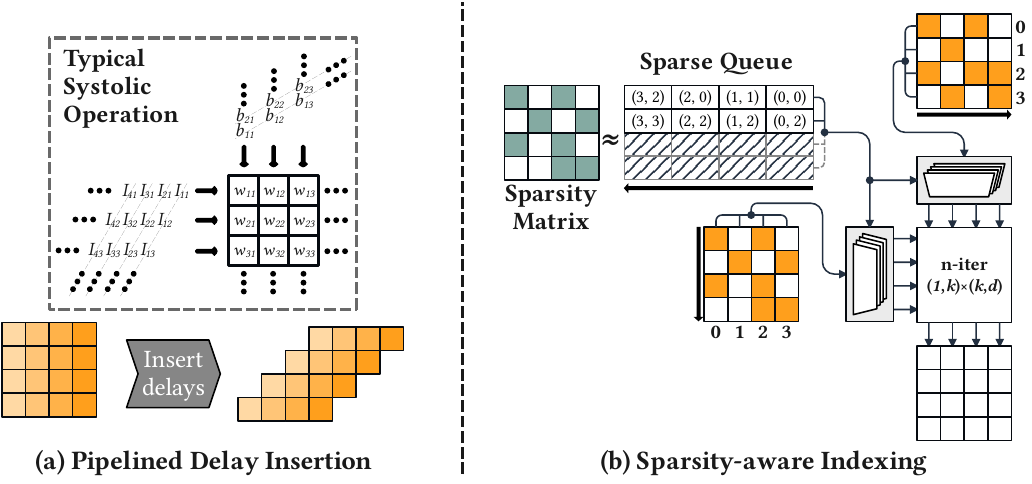}
\vspace{-4mm}
\caption{\textbf{Feed scheduler.} 
(\textbf{a}) Pipelined delay insertion aligns rows/columns for WS/OS systolic waves with no host-side data reshaping.
(\textbf{b}) Sparsity-aware indexed reads via a small BRAM-backed sparse queue (SQB) and address generator feed only active pairs to $1{\times}C_S$ SIMD and RADT, eliminating bubbles and DRAM detours after top-$k$ pruning.}
\label{fig:feed_scheduler}
\vspace{-4mm}
\end{figure}

\subsection{Lightweight Feed and Nonlinear Units}
The feed scheduler in Fig.~\ref{fig:feed_scheduler}(a) inserts per-row delays for correct systolic alignment and, in Fig.~\ref{fig:feed_scheduler}(b), performs sparsity-aware indexed reads using a small BRAM-backed sparse queue, so inputs need no host-side repacking while $1{\times}C_S$ and RADT consume only active pairs. Compact nonlinear units (SoftMax, GELU/ELU, normalization) use polynomial or piecewise-linear approximations and reuse a tree-style accumulate pattern; they are sized to sustain the array’s streaming rate with modest resources.

\subsection{Putting It Together}
At runtime, \design{} applies a simple policy: dense DDMM uses WS or OS according to reuse and width (Fig.~\ref{fig:hw_arch}(c)); SDDMM/SpMM use $1{\times}C_S$ when activity per row is near $C_S$ and uniform, and RADT when activity is much smaller or skewed (also Fig.~\ref{fig:hw_arch}(c)); element-wise stages use normal SIMD; top-$k$ runs in stream so pruning immediately reduces downstream work. This shared-datapath approach maintains utilization across dense and sparse regimes and requires no bitstream reconfiguration.

\section{Software Stack}
\begin{figure}[tb!]
\centering
\includegraphics[width=\linewidth]{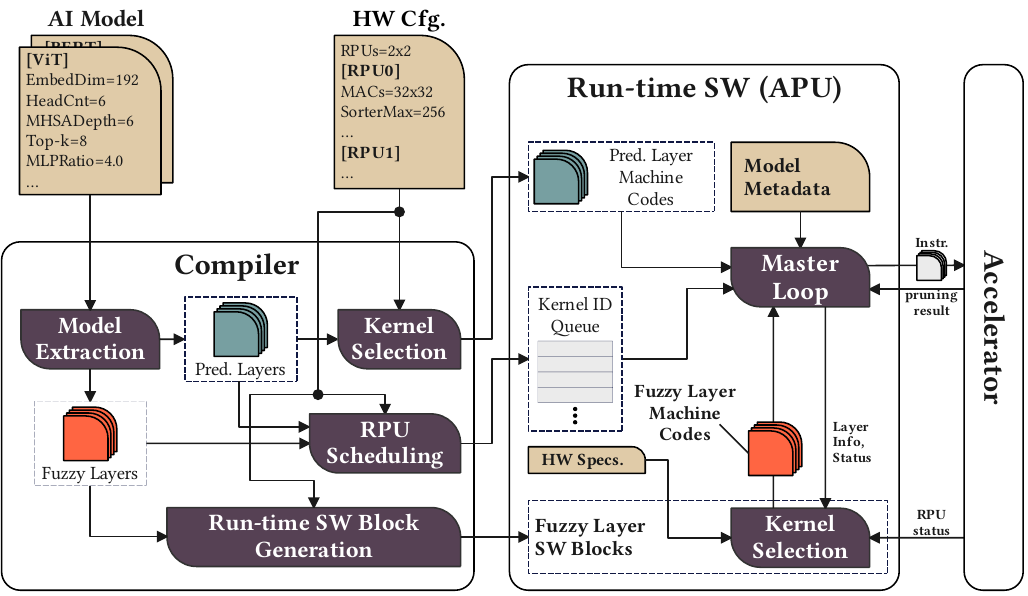}
\caption{\design{} compile–run flow. The compiler parses a structured model, classifies layers (predictable vs. fuzzy), maps them to DDMM/SDDMM/SpMM, selects MSE modes via a sparsity/shape policy, and emits compact instruction blocks plus a dependency DAG. At runtime, an APU-backed controller fills fuzzy templates (e.g., token counts), configures top-$k$, and schedules ready blocks across RPUs using dependency-aware layer offloading (DALO). Pruning indices flow forward to shrink subsequent kernels.}
\label{fig:compile_flow}
\vspace{-4mm}
\end{figure}

\design{} provides a vertically integrated stack that takes a structured model description and produces pruning-aware, runtime-adaptive execution on a single-bitstream FPGA. The stack enables multimodal offloading (vision, language, graph), mixes static and dynamic flows (e.g., variable token counts), and chooses efficient MSE modes per kernel. It comprises a model specification interface, a multi-stage compiler that emits compact instruction blocks, and a lightweight runtime that instantiates fuzzy layers, manages pruning, and dispatches work across the RPU grid (Fig.~\ref{fig:compile_flow}).

\subsection{Compiler Infrastructure}
The model is described in a compact configuration that lists layer types (Attention, MatMul, GELU, SDDMM, etc.), tensor shapes, dataflow hints, and whether token pruning is enabled. During parsing, layers are classified as \emph{predictable} (fixed shapes and static dataflow, e.g., MLPs or static convolutions) or \emph{fuzzy} (runtime-dependent, e.g., variable-length NLP sequences or dynamically pruned attention). This split allows the compiler to generate final binaries for predictable layers and templates for fuzzy ones whose missing fields will be filled at runtime.

Kernels are unified as DDMM, SDDMM, and SpMM and assigned an MSE mode using a simple scoring policy driven by layer shape and expected sparsity: dense kernels map to WS/OS systolic; pruned attention (SDDMM) and message passing (SpMM) map to $1{\times}C_S$ SIMD when activity per row is near the array width and uniform, and to RADT when activity is much smaller or skewed. To keep compile time small, RADT exploration is restricted to a short list of homogeneous partitions consistent with the array dimensions; the compiler evaluates only a handful of candidates rather than a combinatorial space.

The compiler then emits compact instruction blocks that encode mode ID, loop bounds/tiling, buffer and operand IDs, pruning options, and dependency tags. A dependency graph (DAG) across blocks enables dependency-aware layer offloading (DALO): independent kernels (e.g., the $Q$, $K$, and $V$ projections) are placed on different RPUs and overlapped to raise utilization. Static placement respects RPU capabilities (presence of top-$k$, nonlinear units) and co-locates producer–consumer pairs when it reduces memory traffic (e.g., top-$k$ adjacent to its consumer SDDMM/SpMM).

\subsection{Runtime Control and Dynamic Execution}
At runtime, a lightweight APU controller streams instruction blocks and fills in fuzzy templates with observed parameters (token counts, pruning ratios, actual sparsity) before dispatch. Pruning thresholds and $k$-values are programmed into the top-$k$ unit; the resulting indices are propagated forward as compact streams to shrink subsequent SDDMM/SpMM and to guide sparse reads. The controller tracks RPU availability, honors the DAG to preserve dependencies, and opportunistically overlaps independent work across the grid; if observed sparsity deviates from compile-time estimates, it can switch the next block’s MSE mode (e.g., $1{\times}C_S$ to RADT) by updating a small control block without any reconfiguration.

\begin{itemize}[leftmargin=*,nosep]
  \item \textbf{Template instantiation:} complete fuzzy blocks with runtime shapes (e.g., sequence length, kept tokens).
  \item \textbf{Pruning control:} load top-$k$ parameters; consume index streams to gate downstream compute and memory.
  \item \textbf{DALO scheduling:} dispatch ready blocks to idle RPUs, co-locate hot producer–consumer pairs, and maintain backpressure via shared queues.
\end{itemize}

This hybrid compile–run design keeps the critical logic in the compiler (kernel unification, mode scoring, DAG construction) while leaving low-cost decisions (template fill-in, $k$ selection, mode flips under sparsity drift) to the runtime. In practice, this preserves high utilization across dense and sparse regimes, allows pruning to immediately reduce downstream work, and maintains adaptation within a single bitstream.

\section{Evaluation}
We evaluate \design{} on multimodal workloads with latency and power as primary metrics. Experiments are run on two FPGA targets—Alveo U50 and ZCU104—to show scalability, with comparisons to an RTX~4090 and Jetson Orin Nano. We report (i) end-to-end latency on multimodal graphs, (ii) ViT token-pruning impact, and (iii) cross-platform results.

We implement \design{} using a Chisel-based RTL generator~\cite{chisel} and synthesize/place-and-route in Vivado~2022.1. Unless noted otherwise, the Alveo U50 runs at 300\,MHz and the ZCU104 at 200\,MHz; all experiments use \textbf{int8} quantization. The top-$k$ engine supports up to 256 elements (bitonic on U50, dual-layer on ZCU104). \autoref{tab:resource} summarizes the configurations and resource utilization.

\begin{table}[tbh!]
\centering
\caption{Hardware configurations and resource utilization.}
\vspace{-2mm}
\resizebox{\linewidth}{!}{%
\begin{tabular}{l|r|r}
\toprule
                    & \multicolumn{1}{c|}{\textbf{Alveo U50}}   & \multicolumn{1}{c}{\textbf{ZCU104}}           \\
\midrule \midrule
RPU Grid            & \multicolumn{1}{c|}{2×2}           & \multicolumn{1}{c}{1×1}                \\
MSE PE              & \multicolumn{1}{c|}{32×32}         & \multicolumn{1}{c}{32×32}              \\
Top-$k$ Eng.        & \multicolumn{1}{c|}{Bitonic, Max-$k$=256} & \multicolumn{1}{c}{Dual-Layer, Max-$k$=256}   \\
Clock Freq.         & \multicolumn{1}{c|}{300 MHz}              & \multicolumn{1}{c}{200 MHz}                   \\
\midrule
LUT                 & 693,092 / 871,680 (79.51\%)               & 163,112 / 230,400 (70.80\%)                   \\
FFs                 & 379,608 / 1,743,360 (21.77\%)             & 84,064 / 460,800 (18.24\%)                    \\
DSPs                & 4,564 / 5,952 (76.68\%)                   & 1,141 / 1,728 (66.03\%)                       \\
BRAM                & 553 / 1,344 (41.15\%)                     & 139 / 312 (44.55\%)                           \\
URAM                & 78 / 640 (12.19\%)                        & 20 / 96 (20.83\%)                             \\
\midrule
Power               & 20.99\,W                                  & 8.05\,W                                       \\
\bottomrule
\end{tabular}}
\vspace{-2mm}
\label{tab:resource}
\end{table}

Model configurations are listed in \autoref{tab:model_hw_eval}, spanning TinyCLIP (ViT/NLP), MDETR (CNN/NLP), and MissionGNN (ViT/GNN, CNN/GNN). Together, these cover ViT, NLP, GNN, and CNN.

\begin{table}[h]
\centering
\caption{AI model configurations used in evaluation.}
\vspace{-2mm}
\resizebox{\linewidth}{!}{%
\begin{tabular}{l|c|c|c|c} \toprule
\textbf{Model}              & \textbf{Label}  & \textbf{Workload1}  & \textbf{Workload2}    & \textbf{Param. Count (M)}  \\ \midrule \midrule
\multirow{4}{*}{\begin{tabular}[c]{@{}c@{}}TinyCLIP \\ (ViT + NLP)\end{tabular}}   & A               & ViT-8M/16           & TEXT-3M               & 11                \\
                            & B               & ViT-39M/16          & TEXT-19M              & 58                \\
                            & C               & ViT-40M/32          & TEXT-19M              & 59                \\
                            & D               & ViT-61M/32          & TEXT-29M              & 90                \\ \midrule \midrule
\multirow{3}{*}{\begin{tabular}[c]{@{}c@{}}MDETR \\ (CNN + NLP)\end{tabular}}      & E               & ResNet-18           & RoBERTa               & 152               \\
                            & F               & ResNet-50           & RoBERTa               & 166               \\
                            & G               & ResNet-101          & RoBERTa               & 185               \\ \midrule \midrule
\multirow{3}{*}{\begin{tabular}[c]{@{}c@{}}MissionGNN \\ (ViT + GNN)\end{tabular}} & H               & ViT-S/16            & GNN-104               & 6                 \\
                            & I               & ViT-B/32            & GNN-104               & 22                \\
                            & J               & ViT-H/14            & GNN-104               & 632                \\ \midrule \midrule
\multirow{2}{*}{\begin{tabular}[c]{@{}c@{}}MissionGNN \\ (CNN + GNN)\end{tabular}} & K               & ResNet-18           & GNN-104               & 12                \\
                            & L               & ResNet-101          & GNN-104               & 45                \\
\bottomrule
\end{tabular}}
\vspace{-3mm}
\label{tab:model_hw_eval}
\end{table}

\subsection{Multimodal Workload Analyses}
\begin{figure}[tb!]
\centering
\includegraphics[width=\linewidth]{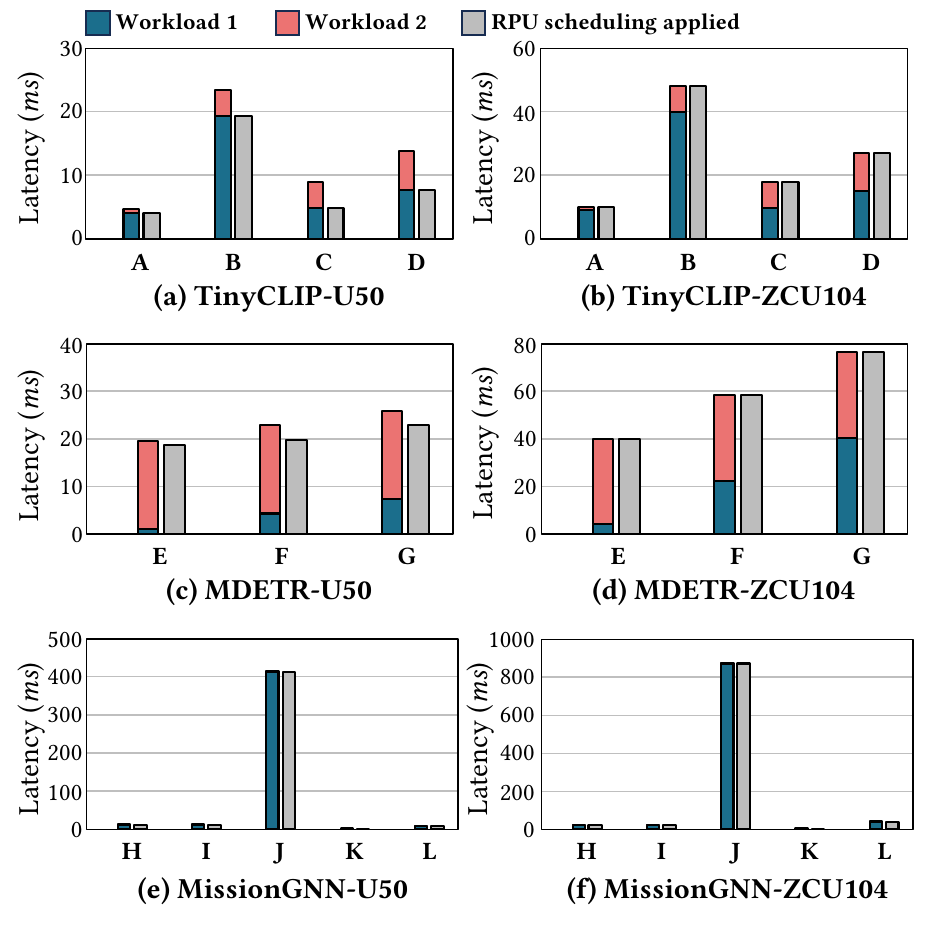}
\vspace{-8mm}
\caption{End-to-end latency with/without DALO; labels refer to \autoref{tab:model_hw_eval}. Token pruning disabled to isolate scheduling effects.}
\label{fig:graph_scheduling}
\vspace{-3mm}
\end{figure}

Fig.~\ref{fig:graph_scheduling} shows the impact of dependency-aware layer offloading (DALO). On U50, TinyCLIP \textbf{A–D} complete in \textbf{4.2–19.4\,ms}; DALO improves throughput by up to \textbf{79.2\%}, keeping latency under \textbf{25\,ms} for interactive rates ($>$40\,FPS). For MDETR \textbf{E–G}, the RoBERTa branch dominates; DALO still yields \textbf{1.0–1.2×} gains by overlapping early text/vision blocks on U50, while single-RPU ZCU104 shows no benefit (sequential execution). MissionGNN \textbf{H–I} finish $<\!$\textbf{12\,ms}; the largest \textbf{J} reaches \textbf{414\,ms} without pruning, indicating pruning is essential for this scale. Overall, DALO converts inter-branch independence into wall-clock savings on multi-RPU devices and degrades gracefully to correct sequential execution on small devices.

\subsection{Token Pruning Impact on Latency}
\begin{figure}[tb!]
\centering
\includegraphics[width=\linewidth]{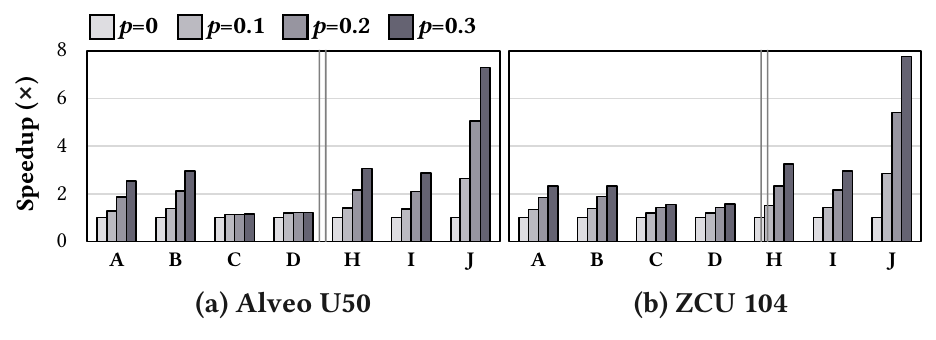}
\vspace{-9mm}
\caption{Speedup from ViT token pruning (DynamicViT); labels per \autoref{tab:model_hw_eval}.}
\vspace{-3mm}
\label{fig:graph_pruning}
\end{figure}
We apply DynamicViT~\cite{dynamicvit} with pruning rates $p\!\in\!\{0.1,0.2,0.3\}$ (baseline $p\!=\!0$), with DALO enabled. Fig.~\ref{fig:graph_pruning} shows up to \textbf{7.3×} (U50) and \textbf{7.8×} (ZCU104) speedups, exceeding GPU-reported gains (\(\sim\)1.6×). Larger-token ViTs (\textbf{A,B,H,J}) benefit most; MissionGNN (\textbf{H–J}) sees higher speedups than TinyCLIP due to ViT’s share of total compute. On ZCU104, pruning yields higher relative gains because scheduling is not the bottleneck. In all cases, pruning immediately reduces downstream SDDMM/SpMM via the in-stream top-$k$ and raises utilization through mode switching.

\begin{figure}[tb!]
\centering
\includegraphics[width=\linewidth]{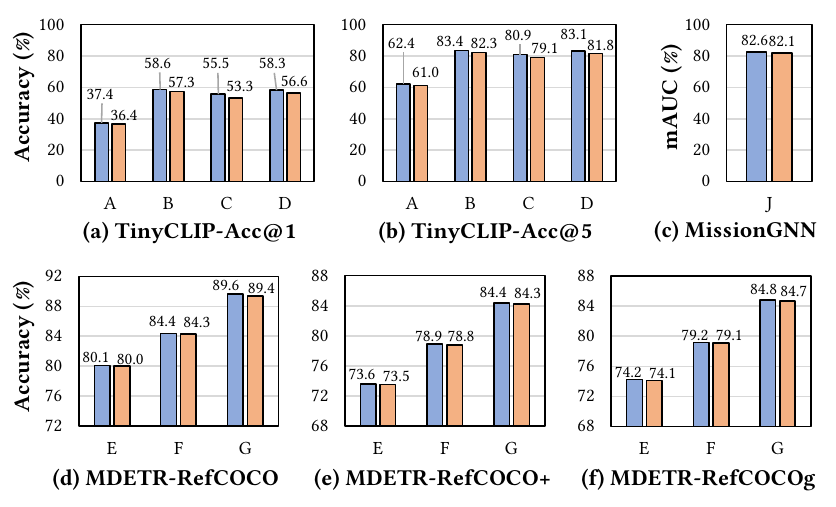}
\vspace{-9mm}
\caption{Accuracy of FP32 vs. int8 quantization; labels per \autoref{tab:model_hw_eval}.}
\label{fig:quantization}
\vspace{-3mm}
\end{figure}

\subsection{Quantization Effects on Accuracy}
As summarized in Fig.~\ref{fig:quantization}, we quantize transformers with Bitsandbytes and use PyTorch static quantization for CNN/GNN. TinyCLIP (\textbf{A–D}) shows \textbf{1.0–2.2\%} top-1 and \textbf{1.0–1.8\%} top-5 drops on ImageNet~\cite{imagenet}; MDETR (\textbf{E–G}) loses \textbf{0.1–0.2\%} on RefCOCO/+/g~\cite{refcoco}; MissionGNN (\textbf{J}) decreases by \textbf{0.5\%} mAUC on UCF-Crime~\cite{ucfcrime}. Across all cases in Fig.~\ref{fig:quantization} (blue: FP32, red: int8), degradation remains under \textbf{2.5\%} while reducing model size by 4× and aligning with \design{}’s int8 datapath.

\begin{table*}[t]
\centering
\caption{Comparison with other FPGA-based accelerators including CNN, GNN, ViT, and NLP transformers.}
\vspace{-2mm}
\resizebox{\linewidth}{!}{%
\begin{tabular}{cc|c||c|c|c|c|c|c|c|c}
\toprule\toprule
\rowcolor{lightblue}&
&
Unit &
\begin{tabular}[c]{@{}c@{}}\textbf{AMD DPU}\\'22 \cite{dpu}\end{tabular} &
\begin{tabular}[c]{@{}c@{}}\textbf{VisionAgile}\\TPDS'24 \cite{hw_cnn_gnn_vit}\end{tabular} &
\begin{tabular}[c]{@{}c@{}}\textbf{GCV-Turbo}\\FCCM'24 \cite{hw_cnn_gnn}\end{tabular} &
\begin{tabular}[c]{@{}c@{}}\textbf{HPTA}\\FPL'23 \cite{hpta}\end{tabular} &
\begin{tabular}[c]{@{}c@{}}\textbf{FTRANS}\\ISLPED'20 \cite{ftrans}\end{tabular} &
\begin{tabular}[c]{@{}c@{}}\textbf{HeatViT}\\HPCA'23 \cite{heatvit}\end{tabular} &
\begin{tabular}[c]{@{}c@{}}\textbf{SWAT}\\ASP-DAC'24 \cite{swat}\end{tabular} &
\begin{tabular}[c]{@{}c@{}}\textbf{TRINE}\\This Work\end{tabular}
\\ \hline

\multicolumn{2}{c|}{ML Workload} &
- &
CNN, Transformer &
CNN, ViT, GNN &
CNN, GNN &
Transformer &
Transformer &
ViT * &
Swin &
\textbf{CNN, GNN, ViT, NLP}
\\

\multicolumn{2}{c|}{Scheduling Mechanism} &
- &
Sequential &
Sequential &
Sequential &
Sequential &
Sequential &
Sequential &
Sequential &
\textbf{Inter-Model Parallel}
\\

\multicolumn{2}{c|}{Number Data Type} &
- &
fixed-point8 &
- &
float16 &
int8 &
fixed-point16 &
fixed-point8 &
int8 &
int8
\\ \hline

\rowcolor{lightblue}\multicolumn{2}{c|}{FPGA Platform} &
- &
ZCU 104 &
Alveo U250 &
Alveo U250 &
ZCU 102 &
VCU118 &
ZCU102 &
Alveo U50 &
Alveo U50
\\

\rowcolor{lightblue}\multicolumn{2}{c|}{Memory Bandwidth} &
GB/s &
19.2 &
77 &
77 &
19.2 &
- &
19.2 &
316 &
316
\\

\rowcolor{lightblue}\multicolumn{2}{c|}{Operating Frequency} &
MHz &
600/300 ** &
600/300 ** &
600/300 ** &
200 &
- &
150 &
200 &
300
\\ \hline

\multicolumn{2}{c|}{Peak Throughput} &
TOPS &
\textbf{2.40} &
- &
1.08 &
- &
- &
0.37 &
0.31 &
1.30
\\

\multicolumn{2}{c|}{DSP Usage} &
- &
- &
\textbf{9091} &
- &
2307 &
6531 &
2307 &
1968 &
4564
\\

\multicolumn{2}{c|}{Maximum Power} &
W &
\textbf{8.85} &
- &
- &
20 &
25.13 &
9.45 &
14.35 &
20.99
\\ \hline

\rowcolor{lightblue}\multirow{3}{*}{\begin{tabular}[c]{@{}c@{}}TinyCLIP\\(ViT + NLP)\\ViT-8M/16+TEXT-3M\end{tabular}} &
Latency &
ms &
- &
4.26 &
- &
6.38 &
9.94 &
7.46 &
22.04 &
\textbf{1.60}
\\

\rowcolor{lightblue} &
Norm. Latency *** &
ms &
- &
7.92 & 
- & 
3.51 & 
10.21 & 
3.55 & 
17.83 & 
\textbf{1.60}
\\

\rowcolor{lightblue} &
Energy & 
mJ & 
- &
- &
- & 
127.54 & 
249.72 & 
70.54 & 
316.27 & 
\textbf{33.58}
\\ \hline

\multirow{3}{*}{\begin{tabular}[c]{@{}c@{}}MDETR\\(CNN + NLP)\\ResNet-101+RoBERTa\end{tabular}} & 
Latency &
ms   &
- & 
30.54 &
- &
- &
- & 
- & 
- & 
\textbf{28.40}
\\
& 
Norm. Latency *** & 
ms &
- & 
58.60 & 
- &
- &
- &
- &
- &
\textbf{28.40}
\\
& 
Energy & 
mJ & 
- & 
- & 
- & 
- & 
- & 
- & 
- & 
596.12 
\\ \hline

\rowcolor{lightblue}\multirow{3}{*}{\begin{tabular}[c]{@{}c@{}}MissionGNN\\(CNN + GNN)\\ResNet-101+GNN-104\end{tabular}} &
Latency &
ms &
21.18 &
9.42 &
8.96 &
- &
- &
- &
- &
\textbf{7.40}
\\

\rowcolor{lightblue}&
Norm. Latency *** &
ms &
9.50 &
17.68 &
16.75 &
- &
- &
- &
- &
\textbf{7.40}
\\

\rowcolor{lightblue}&
Energy &
mJ &
187.44 &
- &
- &
- &
- &
- &
- &
\textbf{155.33}
\\ \hline\hline

\multicolumn{11}{l}{* The original work only supports ViT, but we assume it can support RoBERTa when padding the input sentence tokens with image token size.}\\
\multicolumn{11}{l}{** 600 MHz for computation units and 300 MHz for data buffer.}\\
\multicolumn{11}{l}{*** Normalized the latency based on the utilized DSP count.}

\end{tabular}
}
\label{tab:comp_sota}
\vspace{-3mm}
\end{table*}
\subsection{Cross-Platform Analyses}

\subsubsection{Comparison with FPGA Accelerators}
\autoref{tab:comp_sota} compares \design{} with recent FPGA accelerators across CNN, ViT, GNN, and NLP. Since most baselines are single-modal, we report their strongest per-modal numbers and use FLOPs-scaled estimates when needed (e.g., TinyCLIP on HeatViT~\cite{heatvit}). We also provide DSP-normalized latency to account for device size; this is wall-clock latency divided by the fraction of DSPs used, so designs that achieve more with fewer DSPs are credited appropriately. Unless stated otherwise, all figures are batch size 1, single-sample medians after warm-up, and energy is computed as $E=\text{Power}\times\text{Latency}$.

For TinyCLIP (configuration \textbf{A}), \design{} achieves \textbf{1.6\,ms}, \textbf{2.7}\,$\times$ faster than VisionAGILE~\cite{hw_cnn_gnn_vit} in raw latency and \textbf{2.2}\,$\times$ per-DSP. Energy per inference is \textbf{33.6\,mJ}, a \textbf{2.1}\,$\times$ improvement over HeatViT. The gap comes from in-stream top-$k$ (downstream SDDMM/SpMM immediately shrink) and DALO overlapping $Q/K/V$ with the text branch across RPUs, which raises utilization without extra memory traffic.

For MDETR, \design{} exceeds VisionAGILE by \textbf{1.1}\,$\times$ raw and \textbf{2.1}\,$\times$ per-DSP despite lower external bandwidth. WS/OS mode switching matches layer reuse, and DALO overlaps early CNN and text blocks before fusion. Gains are modest because the unpruned text encoder dominates, but they are consistent across configurations and devices.

For MissionGNN, \design{} is \textbf{1.2}\,$\times$ faster in raw latency and \textbf{1.3}\,$\times$ per-DSP than GCV-Turbo~\cite{hw_cnn_gnn} and AMD DPU~\cite{dpu}, with \textbf{1.2}\,$\times$ better energy efficiency. Native SDDMM/SpMM support in the MSE (switching to $1{\times}C_S$ SIMD or RADT as sparsity skews) sustains utilization on irregular graphs without reconfiguration and avoids host-side packing.

Overall, \design{} is, to our knowledge, the only accelerator that executes CNN, ViT, GNN, and NLP end-to-end on a single bitstream. Competing works often require FPGA reconfiguration (100–150\,ms~\cite{reconfig_latency}) or CPU offload; those costs are typically excluded from reported latencies, so full-pipeline advantages are larger in practice.

\begin{table}[t!]
\centering
\caption{Latency comparison between \design{} and GPUs (ms); ViT uses $p{=}0.3$ configuration.}
\vspace{-2mm}
\resizebox{\linewidth}{!}{%
\begin{tabular}{c|cc|cc|cc||c|cc}
\toprule
\cellcolor{lightblue}\textbf{Workload}
& \multicolumn{2}{c|}{\textbf{TinyCLIP}}
& \multicolumn{2}{c|}{\textbf{MDETR}}
& \multicolumn{2}{c||}{\textbf{MissionGNN}}
& \cellcolor{lightblue}\textbf{Workload}
& \multicolumn{2}{c}{\textbf{TinyCLIP}} \\ \hline
\cellcolor{lightblue}\textbf{Cfg.}                                                       & A     & B     & F     & G     & J     & L     & \cellcolor{lightblue}\textbf{Cfg.}                                                & A            & B             \\ \hline
\cellcolor{lightblue}\begin{tabular}[c]{@{}c@{}}Alveo\\ U50$^\mathrm{*}$\end{tabular}    & 1.6   & 6.5   & 18.7  & 23.0  & 56.9   & 7.4   & \cellcolor{lightblue}ZCU104$^\mathrm{*}$                                                       & 4.4         & 20.8         \\
\cellcolor{lightblue}\begin{tabular}[c]{@{}c@{}}RTX\\ 4090$^\mathrm{*}$\end{tabular}     & 37.1  & 49.9  & 21.7  & 28.4  & 50.1   & 44.1 & \cellcolor{lightblue}\begin{tabular}[c]{@{}c@{}}Jetson\\ Orin Nano$^\mathrm{*}$\end{tabular}   & 30           & 40.6          \\ \hline
\cellcolor{lightblue}\begin{tabular}[c]{@{}c@{}}Speedup\\ (×)\end{tabular}        & \textbf{22.57} & \textbf{7.64}  & \textbf{1.16}  & \textbf{1.24}  & \textbf{0.88}  & \textbf{5.98}  & \cellcolor{lightblue}\begin{tabular}[c]{@{}c@{}}Speedup\\ (×)\end{tabular} & \textbf{6.86}         & \textbf{1.96}          \\
\bottomrule
\multicolumn{10}{l}{Labels refer to \autoref{tab:model_hw_eval}. $^\mathrm{*}$Numbers are latencies (ms).} \\
\end{tabular}}
\label{tab:cross_platform}
\vspace{-3mm}
\end{table}

\subsubsection{GPU Comparison}
\autoref{tab:cross_platform} compares \design{} with an RTX~4090 and a Jetson Orin Nano at batch size 1 and int8 where applicable; ViT uses $p{=}0.3$, and transfers are overlapped when supported.

For TinyCLIP, the Alveo U50 is \textbf{22.57}\,$\times$ faster than the RTX~4090 on configuration \textbf{A} and \textbf{7.64}\,$\times$ on \textbf{B}. On the embedded side, ZCU104 exceeds Orin Nano by \textbf{6.86}\,$\times$ and \textbf{1.96}\,$\times$. GPUs underutilize on irregular sparsity after pruning; the MSE’s sparse-friendly modes and on-chip top-$k$ immediately reduce downstream work.

For MDETR, the U50 delivers \textbf{1.16–1.24}\,$\times$ over the RTX~4090. The text encoder is unpruned and CNNs map well to tensor cores, so the margin is smaller.

For MissionGNN, configuration \textbf{L} achieves \textbf{5.98}\,$\times$ versus RTX~4090 due to efficient SpMM, while the very large ViT in configuration \textbf{J} yields \textbf{0.88}\,$\times$ (limited by FPGA resources rather than dataflow). Even here, performance per watt remains favorable: FPGA boards operate at or below \textbf{21\,W}, whereas the RTX~4090 draws roughly \textbf{430\,W}.

\section{Conclusion}
We presented \design{}, a single-bitstream, runtime-adaptive accelerator and compiler that unifies ViT/CNN/GNN/NLP as DDMM/SDDMM/SpMM on a shared mode-switchable PE array; the MSE switches among WS/OS, $1{\times}C_S$ SIMD, and RADT, while a width-matched in-stream top-$k$ and DALO sustain utilization without reconfiguration. Across TinyCLIP, MDETR, and MissionGNN on Alveo U50 and ZCU104, \design{} cuts latency by up to $22.57\times$ and $6.86\times$ (vs.\ RTX~4090 and Orin Nano, respectively) at $\le$21\,W; pruning alone yields up to $7.8\times$ on ViT-heavy cases and DALO adds up to $79\%$ throughput, with $<2.5\%$ accuracy loss under int8. Future work includes extending run-time sparsity to NLP/CNN, exploring heterogeneous RADT partitions with fast cost models, and finer-grained kernel splitting with telemetry-guided policies to push \design{} toward a broadly deployable, energy-efficient substrate for multimodal AI.

\bibliographystyle{ACM-Reference-Format}
\bibliography{IEEEabrv, References/intro, References/rw_multimodal_ai, References/rw_sparse_vit, References/rw_hw_arch, References/technical_contents}

\end{document}